# Attentional modulation of visual spatial integration: psychophysical evidence supported by population coding modeling[1]

Authors: Alessandro Grillini, Remco J. Renken, Frans W. Cornelissen

[1] Laboratory of Experimental Ophthalmology, University Medical Center Groningen, University of Groningen, The Netherlands

[2] Cognitive Neuroscience Center, University Medical Center Groningen, University of Groningen, The Netherlands

Corresponding author:
    Alessandro Grillini
Contact information:
    University Medical Center Groningen, P.O. Box 196, 9700 AD, Groningen, The Netherlands
    a.grillini@rug.nl

# Abstract

Two prominent strategies that the human visual system uses to reduce incoming information are spatial integration and selective attention. While spatial integration summarizes and combines information over the visual field, selective attention can single it out for scrutiny. The way in which these well-known mechanisms – with rather opposing effects – interact remains largely unknown. To address this, we had observers perform a gaze-contingent search task that nudged them to deploy either spatial or feature-based attention to maximize performance. We found that, depending on the type of attention employed, visual spatial integration strength changed either in a strong and localized or a more modest and global manner compared to a baseline condition. Population code modeling revealed that a single mechanism can account for both observations: attention acts beyond the neuronal encoding stage to tune the spatial integration weights of neural populations. Our study shows how attention and integration interact to optimize the information flow through the brain.

# Introduction

Perception is a complex task. Our sensory systems constantly deal with large amounts of incoming information that, in order to be usable, must be reduced and selected. In vision, two relevant mechanisms for this purpose are spatial integration and selective attention. In our context, with the term "integration", we specifically refer to the summation of information coming from different regions of the visual field. Visual spatial integration is a prominent process throughout the visual system, evident from retinal ganglion cells integrating the output of bipolar cells (Barlow, 1953; Freed & Sterling, 1988; Hartline, 1940b) to the increasingly larger receptive fields along the cortical visual hierarchy (Smith, Singh, Williams, & Greenlee, 2001). Importantly, visual spatial integration should not to be confounded with other proposed types of "integration", such as in Feature Integration Theory (Treisman & Gelade, 1980), where the integration concerns the combining of different features in the same spatial region.

Visual spatial integration can assume different forms and functions. For example, an advantageous form of spatial integration is contour detection (Field, Hayes, & Hess, 1993), while a - seemingly - disruptive one is crowding; i.e. the jumbled perception of objects surrounded by nearby distractors when these are viewed with peripheral vision (Bouma, 1970). Both contour integration and crowding have their roots in the Gestalt principles and while being rather different phenomena, previous studies suggested that they may be different manifestations of the same underlying integrative process and share common "association fields" (Chakravarthi & Pelli, 2011; Tomer Livne & Sagi, 2007; T. Livne & Sagi, 2010; May & Hess, 2007). In this study, we will focus on the crowding phenomenon to study integration as it is relatively straightforward to precisely measure and by manipulating the presence/absence of distractors or their properties (for a review see (Levi, 2008)).

While crowding is usually presented as a limitation of vision, recent computational studies suggest that it actually supports optimal visual decision-making given certain constraints of the underlying neural mechanisms (Dayan & Solomon, 2010; Sun, Chung, & Tjan, 2010; van den Berg, Johnson, Martinez Anton, Schepers, & Cornelissen, 2012). Various efforts to model the mechanisms underlying crowding and visual integration (Harrison & Bex, 2015; Nandy & Tjan, 2012; van den Berg, Roerdink, & Cornelissen, 2010) have turned to population coding principles (Pouget, Dayan, & Zemel, 2000). In this approach, the integral activity of populations of neurons encodes the probability distribution of a specific feature (for instance, orientation) at a certain spatial location. The success of these models in explaining various properties of crowding would imply that it is primarily "hard-wired", feedforward and compulsory in nature and simply a consequence of an overlap of large receptive fields in the visual periphery (Greenwood, Bex, & Dakin, 2010; Parkes, Lund, Angelucci, Solomon, & Morgan, 2001). However, at the same time, a number of studies suggest that crowding can be modulated, either by perceptual learning (Zhu, Fan, & Fang, 2016) or by attention (Bacigalupo & Luck, 2015; He, Cavanagh, & Intriligator, 1996a; Mareschal, Morgan, & Solomon, 2010; Põder, 2006, 2007). The current generation of integration models does not accommodate such modulations, indicating a gap in our understanding of the phenomenon.

Recently, the effect of visual attention on modulating the responses of neural populations has been described in studies of animal models (Kanashiro, Ocker, Cohen, & Doiron, 2017; Rabinowitz, Goris, Cohen, & Simoncelli, 2015) and several studies relying on population coding models provide a theoretical framework to explain attentional modulations (Cohen & Maunsell, 2011; Herrmann, Heeger, & Carrasco, 2012; Ling, Liu, & Carrasco, 2009; Pestilli, Ling, & Carrasco, 2009).

Hence, despite the fact that both attention and integration are continuously employed in visual perception, a formal characterization of their interaction is still lacking. Having this would significantly enhance our understanding of how cognition operates in concert with early vision to flexibly adapt it to the on-demand information needs of different tasks.

Here, we address this question by asking observers to perform a search task designed in such a way that a change in attentional deployment is necessary to optimize their performance. We use two distinct constraints: one that removes all foveal visual information and one that removes only the task-relevant information from foveal vision. In the former case, observers have to search for the target in their periphery beyond the edge of the visual constraint, i.e. enhance spatial attention (SA). In the latter case, when only the orientation information is removed and no spatial cues are given, observers have to find the target by focusing on subtle changes in orientation that can happen anywhere on the visual field, i.e. enhance feature-based attention (FBA). Before and after these attentional manipulations we measure spatial integration strength, operationalized as the crowding effect (i.e. the difference in performance between discrimination of targets presented in isolation and discrimination of targets surrounded by distractors). In addition, we assess eye-movement and pupil dilation as indices of attentional engagement (Kahneman & Beatty, 1966).

To preview our main finding, our results corroborate that selective attention changes the strength of visual spatial integration, yet with different patterns depending on whether all foveal information was removed or only task-relevant information. Our result is specific to integration and does not reflect a general change in sensitivity as we found no evidence for similar attentional modulation for targets presented in isolation (such that integration cannot occur). We show how population coding can coherently account for these different patterns: selective attention acts beyond the encoding stage of vision and changes the relative contributions of neurons to the population responses underlying integration. Our work provides a unified and mechanistic account for how different types of selective attention are able to modulate visual spatial integration in the human visual system.

# Methods

## *Observers*

A total of 10 healthy participants took part in the experiments (age range 19 - 29, 3 females, 7 males). One was an author (AG), while the remaining participants were student-volunteers, naïve to the purpose of the study. All participants had normal or corrected to normal vision, which was verified prior to data collection (Freiburg Vision Test 'FrACT', (Bach, 2007). Participants received either a small financial reward or course credits for their participation in the study. The study followed the tenets of the Declaration of Helsinki. The ethics board of the Psychology Department of the University of Groningen approved the study protocol. All participants provided written informed consent prior to participation.

## *Materials*

The experiment was designed and conducted using MATLAB extended with Psychtoolbox-3 (Brainard, 1997; D. G. Pelli, 1997) and Eyelink Toolbox (Cornelissen, Peters, & Palmer, 2002). Stimuli were displayed on a 22-in LaCie CRT monitor with a refresh rate of 120 Hz. Gaze position and pupil diameter were monitored and recorded at 500 Hz with an Eyelink 1000 (SR Research, Kanata, Ontario, Canada) infrared eye-tracker at a viewing distance of 60 cm. The Eyelink's built-in 9-point procedure was used for calibration prior to each run. A head-chin rest was used to minimize head movements.

## *Stimuli and Procedure*

The experiment was conducted in a dark and quiet room in three sessions of two hours each. To limit fatigue for the observers, these sessions were conducted over three consecutive days.

Each session consisted of two alternating tasks: a visual search to induce attentional engagement and a 2-alternative forced choice (2-AFC) orientation discrimination task with crowded and isolated targets to measure spatial integration. Starting with a visual search period of 180 seconds, we then proceeded to alternate each trial of the 2-AFC task with 5 seconds of visual search to ensure continuous attentional engagement. A scheme of the experimental session is shown in Figure 1-A.

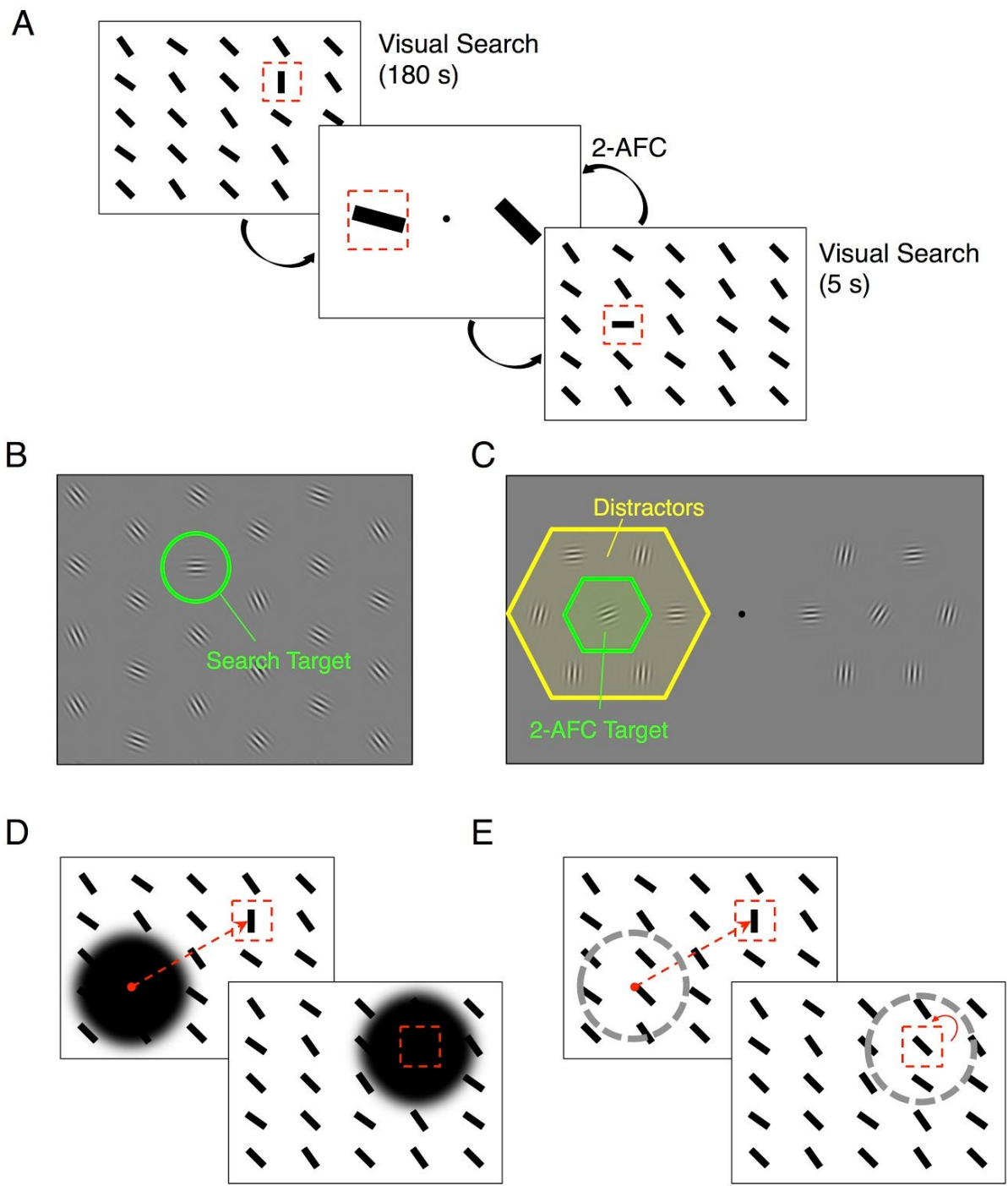

**Figure 1.** Experimental stimuli and procedure. **A**, Schematic representation of the experiment timeline: the first 180 seconds of visual search are followed by an alternance between one 2-AFC trial and 5 seconds of visual search, for a total of 120 trials. This structure ensures a constant and sustained attentional engagement. **B**, Example of the grid of

Gabor patches used during the visual search task. The hexagonal pattern serves the purpose of assuring equidistance between any adjacent pair of Gabor patches. The target is identified as the only one in the grid being perfectly horizontal or vertical. **C**, Example of 2-AFC stimuli. The hexagonal pattern of distractors is preserved, while the targets are shown at different eccentricities from the fixation point. The observers must identify the target that is closest to horizontal orientation. **D**, Schematic representation of the effect of the *visual deprivation*: the mask follows the observers' gaze completely occluding the stimuli. In order to visualize the target the observers must keep their gaze away from it. **E**, Schematic representation of the effect of the *information deprivation*: the "mask" is invisible to the observers, but they are instructed that if their gaze falls too close to the target, this would change its orientation, becoming indistinguishable from the distractors. In this way, no spatial cue is provided, the visual stimulation is preserved but any task-relevant information is removed from the fovea.

## *Visual search*

In the visual search task, the observer had to localize a target (the only one either vertical or horizontal) amongst a grid of tilted distractors. Both target and distractors consisted of Gabor patches (diameter = 1 deg of visual angle; 50% contrast; spatial frequency = 6 cycles-per-degree) placed in a hexagonal grating with a center-to-center distance of 2 degrees of visual angle. All the patches were shifting in phase at a rate of 1 Hz in random directions in order to provide a constant stimulation and to minimize peripheral filling-in. A sample search grid is shown in Figure 1-B.

In order to nudge the observers to shift their attention towards the periphery, two constraints were tested: complete removal of foveal input (*visual deprivation*) and removal of task-relevant information from the foveal region (*information deprivation*). A regular visual search without any constraint served as control condition (*baseline*). The order of presentation of the conditions was randomized for each observer. Both constraints were implemented as on-screen circular masks coupled in real time with the gaze of the observers. The delay between gaze acquisition and display of the mask on screen was below 10 ms (approx. 1 frame with a refresh rate of 120 Hz). In both conditions, the radius of the mask (5.5 degrees) was more than twice the center-to-center distance of the Gabors, such that – according to Bouma's law

(Bouma, 1970) – the target perceived in the periphery would always be crowded by the surrounding distractors.

In the *visual deprivation* condition, the mask completely prevented the observers from seeing the target (schematic representation in Figure 1-D). In the *information deprivation* condition, the mask consisted of an invisible circular area that changed the orientation of a target falling within its boundaries, so that it was no longer distinguishable from the distractors (schematic representation in Figure 1-E). The purpose of these constraint was to enforce an attentional shift from the center to the periphery of the visual field, but in two distinct manners: while the edge of *visual deprivation* clearly indicated where the observers should focus (spatial attention - SA), the *information deprivation* did not provide any spatial cue, forcing the observers to focus (peripherally) on the feature of interest (orientation) rather than a specific location (feature-based attention - FBA). During the constrained search tasks the observers were instructed to "search with the corner of their eye" and to press a key corresponding to the perceived orientation the moment they located the target (left arrow key = vertical target, right arrow key = horizontal target).

## *2-Alternative Forced Choice*

The 2-AFC test served to quantify spatial integration by measuring the strength of visual crowding effect. The observers were instructed to fixate a dot in the center of the screen and to judge the orientation of two targets presented peripherally on both sides. Their task was to decide which of the two was oriented most horizontally and press a key accordingly (left/right arrows). To measure visual crowding, we presented the targets both in isolation and surrounded by distractors. Targets and distractors were Gabor patches identical to those used in the search task, placed in the same hexagonal pattern (see Figure 1-C and compare to Figure 1-B). Stimuli were displayed for 300 ms with no time constraint on the observer response. To measure the effect of integration in different locations of the visual field, we show the targets at four possible different eccentricities from the fovea: 3, 5, 7, and 9 degrees of visual angle. In this way we sampled locations both within, on the edge and outside the visual constraints present during the visual search. Each observer completed a total of 1440 trials (120 trials ✕ 4 eccentricities ✕ 3 conditions, the order between the eccentricities and conditions was randomized for each observer). The 2-AFC task was repeated without variations across the three visual search task conditions (*baseline*, *visual deprivation*, *information deprivation*). In the isolated condition, a low contrast grey circle was displayed surrounding the targets. That is to provide a spatial cue for the location of the targets that in the flanked condition is provided by the distractors themselves (Petrov, Verghese, & McKee, 2006). To ensure central fixation and to acquire pupillometric data, each trial would start only after the observers fixated on the fixation dot for 1 second.

## *Statistical Analysis*

All data was analyzed using custom-made scripts and built-in functions of MATLAB.

In the visual search task, a target detection location was defined by the *x y* screen coordinates of the last fixation made before pressing the button response, acquired with the eye-tracker. Each location was normalized with respect to the target position and expressed in terms of eccentric distance. The preferred detection eccentricity for each condition was obtained by the peak of the probability density distribution of detection eccentricities, as determined by fitting it with a lognormal function (illustrated in Figure 4-B).

The pupillometric analysis was performed by segmenting the data in two epochs: a time window of ± 2 seconds around the instant of each target detection and another time window of 1 second during the fixation validation prior to each 2-AFC task. To remove spikes in the pupil measurement caused by blinks and eye-movements (during the search task), each epoch was processed with a third-order one-dimensional median filter with a time window of 50 ms. For each condition (*baseline*, *visual deprivation*, *information deprivation*) we computed its grand mean as the average between the epochs of all subjects and all eccentricities.

In the 2-AFC experiment, both for single observer and group analyses, psychometric curves were obtained by fitting the data with a logistic regression model. The Just Noticeable Difference (JND) threshold was computed as the difference in the orientation of the target necessary to move from the 50% to the 75% correct point on the psychometric curve (as illustrated in Figure 3-A). The 95% confidence intervals for each JND were obtained with a standard bootstrap resampling procedure (10,000 repetitions).

Integration strength was calculated as the $JND_{flanked}$ - $JND_{isolated}$ for each separate pair of conditions (eccentricity × search task condition). For the group analysis, the resulting integration strengths were averaged between observers and then compared across the three different conditions (*baseline*, *visual deprivation*, *information deprivation*) grouped by eccentricity.

We analyzed the presence of feature-level modulations, i.e. whether the feature-based attention had an affect on the sensitivity towards the *level* of our feature of interest (i.e. horizontal orientation vs vertical orientation) besides the feature *type* (i.e. orientation vs contrast). To do so, we investigated whether repeatedly searching for a specific orientation (horizontal and vertical) led to a change in performance in the 2-AFC task, compared to non-searched orientations. We grouped the 2-AFC trials based on their targets' orientation, separating those either horizontal or vertical (the target orientations in the visual search task) from all the other orientations. We considered as horizontal any target between 0° and 15° while we considered vertical any target between 75° and 90°. Then, we computed the performance for each orientation bin in terms of percentage of correct responses. Finally, we calculated the differences between *baseline* and either *visual* or *information deprivation*, both for isolated and crowded targets. We performed two-tail two-sample t-tests ($\alpha = 0.05$) to evaluate the differences between "search" orientations and "other" orientations.

Unless stated otherwise, all other statistical comparisons were performed with one-tailed nonparametric permutation testing (number of permutations = 1024) at 0.05 significance level. Family Wise Error correction was applied to all comparisons by tracking the max statistic across the three conditions for each permutation. A significance threshold was set as the 100-α percentile of the thus obtained distributions.

# Results

Our main result is that visual integration strength is reduced after constraining the observers to attend to their visual periphery to perform a search task, compared to a baseline condition with no constraint (see Figure 4-A). However, we find that the pattern of reduction depends on the type of visual constraint applied during the search task. In the *visual deprivation* condition, the reduction in visual integration strength compared to the *baseline* was spatially selective and showed a strict correspondence to the preferential retinal eccentricity also used to detect the target during the search task. However, in the selective absence of only the task-related information during search (*information deprivation* condition*)*, the reduction was more modest and spatially non-selective – even though the preferential retinal eccentricity for target detection did not change. Isolated target discrimination was not affected by any of our manipulations. Below, we describe these results in detail.

Figure 2 shows the spatial distribution of target detection locations during the search task. In the *baseline* condition, the average location corresponds to the foveal region and the spread is relatively small (median = 1.10±0.72 degrees of visual angle) while in both *deprivation* conditions the preferred locations lie outside the scotoma boundary and show more dispersion (*visual deprivation* median = 7.27±1.97 deg; *information deprivation* median = 7.99±1.55).

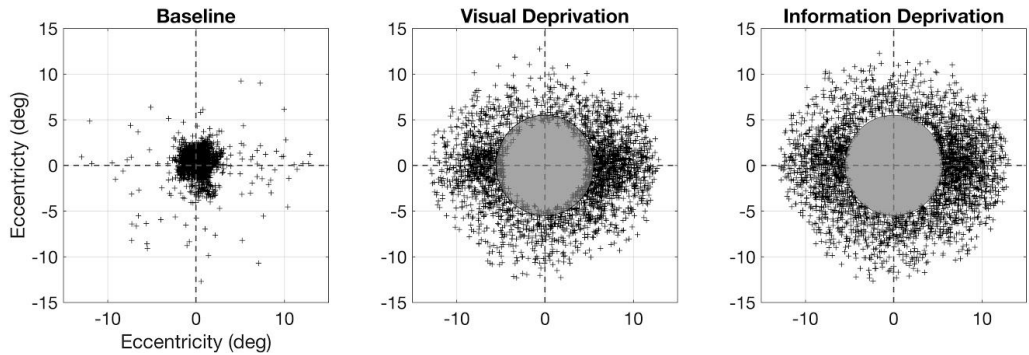

**Figure 2.** Results of the visual search task to induce attentional modulations. Each small black cross represents the last fixation made just before each target detection response. The location of the fixation in the visual field has been normalized with respect to the position of the target, represented by the intersection of the dashed lines at the center of the visual field. The superimposed circular gray areas represent the deprived region in the conditions *visual deprivation* and *information deprivation*.

For isolated targets, in all conditions, JND thresholds are approx. similar at all eccentricities and results do not differ significantly amongst conditions ($p_{(baseline\ vs.\ visual\ deprivation)}= 0.808$; $p_{(baseline\ vs.\ information\ deprivation)}= 0.211$, two-tailed permutation tests). However, for crowded targets, the results depend on the condition (Figure 3-B).

In the *baseline* condition we find that the orientation discrimination thresholds increase linearly with the eccentricity of the stimuli, thus replicating the classic 'crowding effect' (Bouma, 1970). In the *visual deprivation* condition, this monotonic increase is absent. Instead, we observe a selective reduction in thresholds for targets shown at 7 deg of eccentricity (Figure 3-B, left panel). In the *information deprivation* condition, we find a reduction in thresholds that is similar at all tested eccentricities (Figure 3-B, right panel).

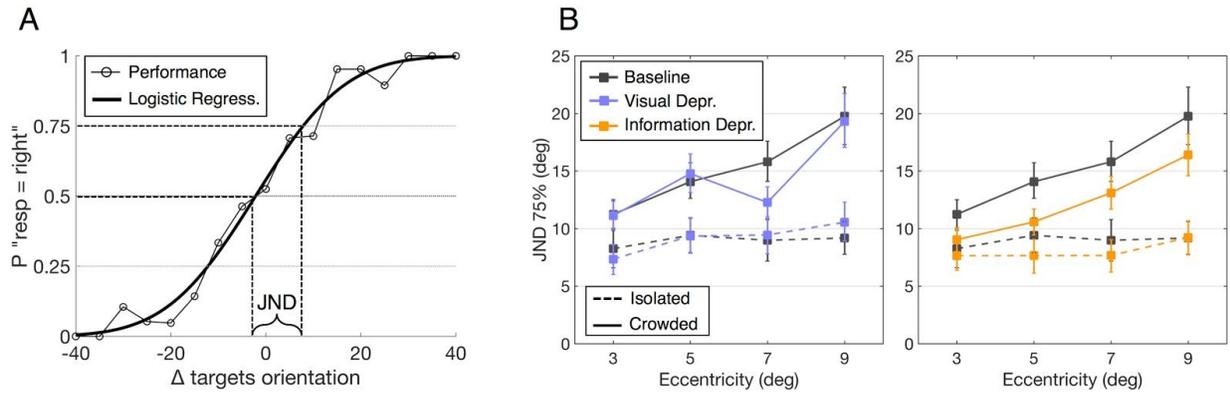

**Figure 3.** Results of the 2-AFC orientation discrimination task to measure integration strength. **A**, Example of psychometric curve. The dots mark the observer's performance while the line represents the logistic regression model that best approximates the data. The Just Noticeable Difference (JND) is computed as the difference between the projections on the x axis coming from 75% and 50% points on the logistic regression fit. **B**, Group results of the 2-AFC task. Each square represents the average JND threshold necessary to achieve 75% correct performance at that specific eccentric location. Lower JND values indicate higher sensitivity and better performance. Error bars represent the 95% confidence intervals obtained with a bootstrap resampling procedure.

Figure 4-A shows changes in orientation discrimination thresholds expressed as integration strength ($JND_{crowded} - JND_{isolated}$) over eccentricities for the conditions tested. Both *deprivation* conditions result in a modulation of integration, but in distinct ways.

In the *visual deprivation* condition, there is a strong reduction specific to the 7 deg target location ($p_{(baseline\ vs.\ visual\ deprivation)} = 0.008$). This location corresponds quite closely to the preferential retinal location used for target detection (6.5 degrees; Figure 4-B). In the *information deprivation* condition, we find a more modest but still significant reduction in integration strength ($p_{(baseline\ vs.\ information\ deprivation)} = 0.027$) that is approx. similar for all target locations. The average preferential retinal location used for target detection is very similar to the other scotoma condition (6.25 degrees; Figure 4-B).

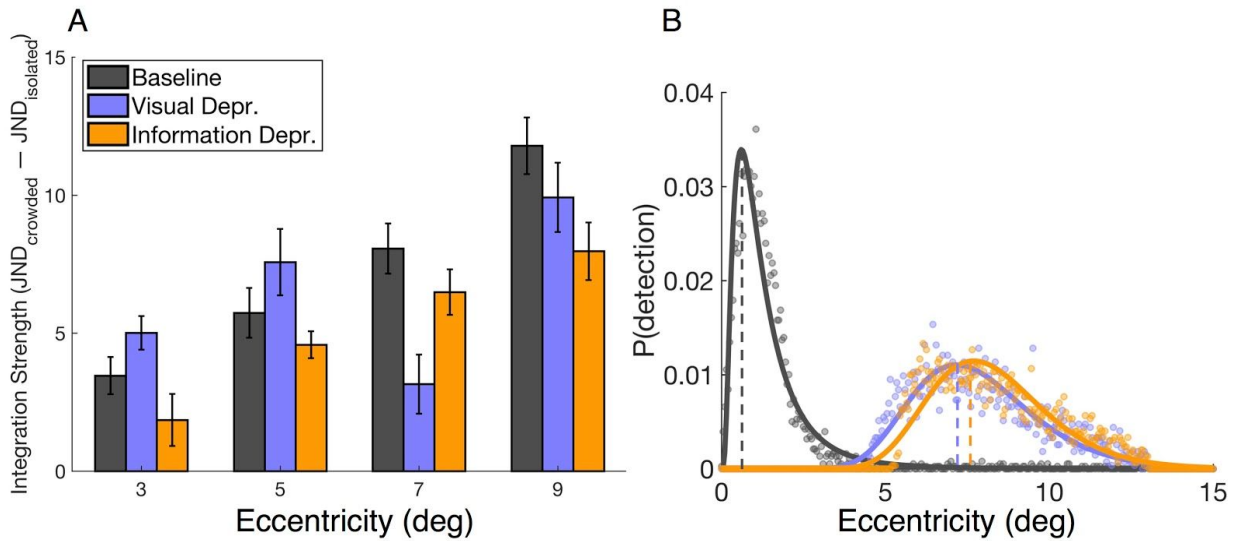

**Figure 4. A**, Integration strength as a function of eccentricity. Error bars represent the standard deviation from the group average **B**, Probability density distributions of detection locations. The detection location eccentricities are shown in bins of 0.25 degrees of visual field. The preferential detection location is the eccentricity corresponding to the peak value of the lognormal fit to the empirical probability density distribution of detection locations.

Figure 5-A shows the results of the pupillometry for the visual search task. There is a significant increase ($p_{(baseline\ vs.\ information\ deprivation)}$ = 0.023) of the absolute pupil diameter during the *information deprivation* condition compared to the *baseline*, but when tested the local variation (expressed as a percentage of change relative to the mean diameter during each epoch) we found no significant increase ($p = 0.943$). The *visual deprivation* condition consistently showed a decrease in pupil diameter compared to the *baseline* one, but this difference is not statistically significant ($p_{(baseline\ vs.\ visual\ deprivation)}$ = 0.052, two-tailed permutation test). During the fixation period of the 2-AFC task we did not find any significant difference in pupil diameter ($p_{(baseline\ vs.\ visual\ deprivation)}$ = 0.300; $p_{(baseline\ vs.\ information\ deprivation)}$ = 0.305), two-tailed permutation test) (Figure 5-B).

The only statistically significant result (Figure 5-A, *information deprivation vs. baseline*) indicates that the attentional modulation occurs in a sustained way during the search task rather than in a transient manner at the instant of target localization. This effect does not carry over into the 2-AFC task.

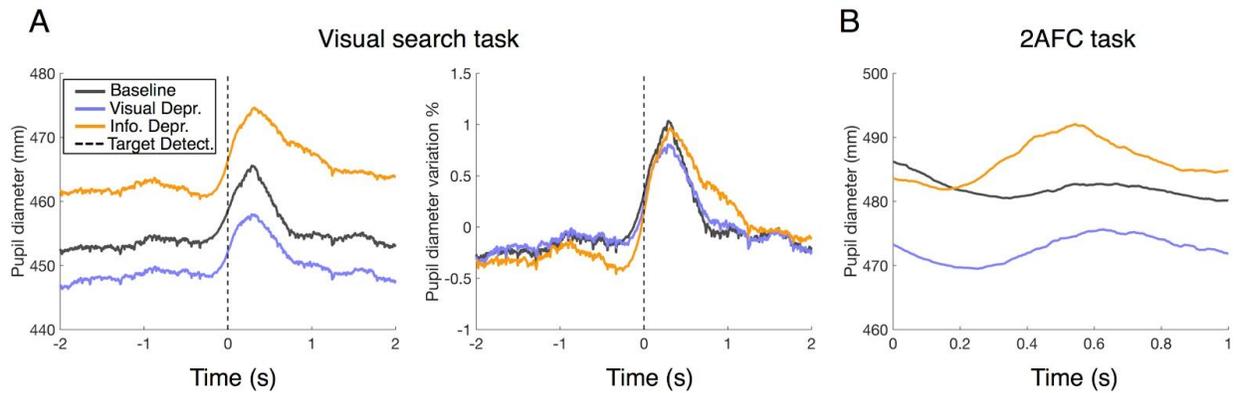

**Figure 5.** Results of the pupillometric analysis to measure attentional engagement across the different tasks. **A**, Grand mean of pupil diameter during the epoch corresponding to ±2 seconds around the instant of each target's detection (left side) and grand mean of relative pupil diameter variation (right side). **B**, Grand mean of pupil diameter during the epoch corresponding to 1 second after the start of each 2-AFC trial (fixation check).

Figure 6 shows the results of the feature-level analysis for the 2-AFC task. Panel A shows the absolute performances for the three conditions separately, while panel B shows their relative differences (*baseline* vs. *visual deprivation* and *baseline* vs. *information deprivation*). Table 1 shows the statistics for the comparisons shown in Panel A, while table 2 shows the statistics for the comparisons shown in Panel B.

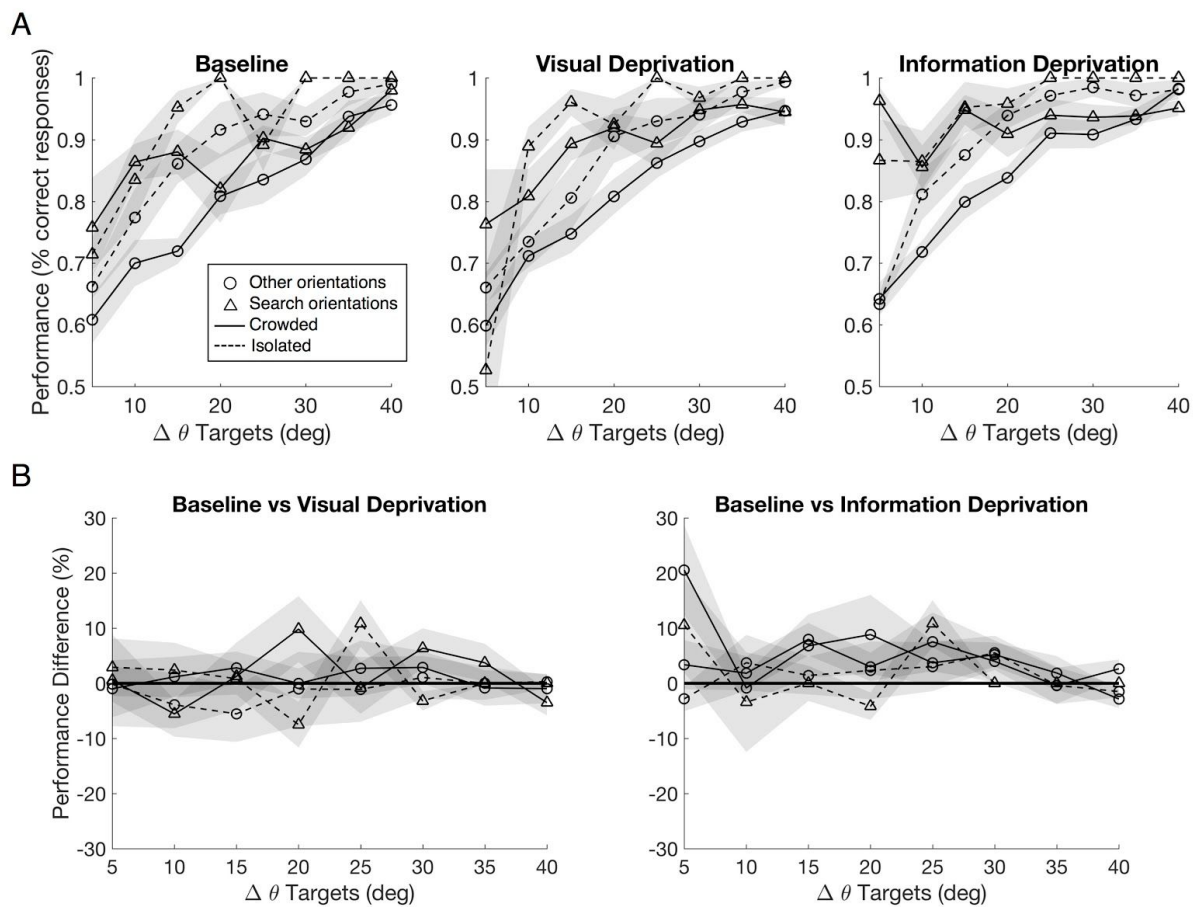

**Figure 6**. Results of the feature-level analysis of the 2-AFC task to assess the nature of feature-based attentional modulations. **A**, Percentage of correct answers in function of Δθ between targets, divided in *isolated* vs *crowded* (dashed and full lines) and *search orientations vs other orientations* (triangles and circles). The shaded areas corresponds to ± 1 SEM. *Search orientations*: $0° \leq \theta \leq 15°$ (horizontal) & $75° \leq \theta \leq 90°$ (vertical); *Other orientations*: $15° < \theta < 75°$. **B**, Differences in performance (expressed in percentage of change) between *baseline vs visual deprivation* and *baseline vs information deprivation*. Baseline is represented by the thick horizontal straight line.

**Table 1. Statistics for feature-level analysis - absolute performances**

|  |  | "Search" vs "Other" orientations (absolute performances) | | |
|---|---|---|---|---|
|  |  | **Baseline** | **Visual Depriv.** | **Information Depriv.** |
| **Isolated** | **T-stat (df = 134)** | 2.3342 | 2.4141 | 3.1431 |
|  | **p-value** | 0.0211 | 0.0171 | 0.0020 |
| **Crowded** | **T-stat (df = 134)** | 3.4029 | 3.9302 | 4.463 |
|  | **p-value** | 0.0008 | 0.0001 | < 0.0001 |

**Table 2. Statistics for feature-level analysis - difference in performances**

|  |  | "Search" vs "Other" orientations (difference in performances) | |
|---|---|---|---|
|  |  | **Baseline vs Visual Depriv.** | **Baseline vs Information Depriv.** |
| **Isolated** | **T-stat (df = 134)** | 0.7126 | -0.2355 |
|  | **p-value** | 0.4773 | 0.8142 |
| **Crowded** | **T-stat (df = 134)** | 0.3387 | 0.3050 |
|  | **p-value** | 0.7353 | 0.7608 |

In all the conditions tested we observe a significant change in absolute performance between "search" and "other" orientation, both for *isolated* and *crowded* trials (Table 1). However, none of these changes remains significant when the comparison is done between the differences in performance following the *deprivation* conditions (Table 2). These results indicate that the differences depending on orientation level are not induced by the attentional modulations.

# Modeling

Population models can explain spatial integration as an overlap in activation of adjacent receptive fields when two stimuli are sufficiently close. The noise distributions of stimuli's perceived positions are well represented by Gaussian distributions of which standard deviations scale with eccentricity, thus leading to stronger crowding in the periphery (Michel & Geisler, 2011). To test whether our observed changes in spatial integration following attentional modulations are consistent with these notions, we modified and extended a previous model (van den Berg et al., 2010) based on population coding principles (Pouget et al., 2000)) by enabling attention to modify the spatial weighting of the contribution of individual neurons to the population response. The purpose of our model is to understand how visual information is encoded and integrated at a neural population level under different attentional states. This can provide valuable insight about the neural mechanism underlying the observed behavior. We chose the population coding approach for a number of reasons. First, it allows a formal description for all the stages of visual information processing (encoding, integration, decoding). Second, it is relevant and meaningful in the context of both spatial and feature-based attention (Cohen & Maunsell, 2011) it allows an easy computation of neural noise correlation, that we are going to use to quantify the performance of the model. Neural noise is an intrinsic property of neuronal activity and it is generally shared among neurons within the same population. The noise correlation within a single population and across the repeated presentation of identical stimuli has been shown to be strongly related with psychophysical performance (Zohary, Shadlen, & Newsome, 1994) and, in general, the amount of noise correlation indicates how much information is encoded by a neural population (more correlation = less information) (Abbott & Dayan, 1999; Averbeck, Latham, & Pouget, 2006; Nirenberg & Latham, 2003). In our context these correlations are particularly relevant, as they have been shown to be an effect of attentional modulations:

attention improves performance by reducing interneuronal correlations (Cohen & Maunsell, 2009; Mitchell, Sundberg, & Reynolds, 2009).

In our model the stimulus at different locations is encoded (Layer I and II) as the average neural population firing rate over the feature space with an added Poisson-like noise component. The average firing rate is described by Equation 1, while Equation 2 describes the population encoding with noise.

Eq. 1:

$$f_i(s) = g \cdot e^{-\frac{(s-s_{pref})^2}{2\sigma_t^2}} \quad , \quad -\pi \leq s < \pi$$

$s - s_{pref}$ is the angular difference between the stimulus orientation space and the preferred orientation of a neural population, $g$ is a gain factor and $\sigma_t$ is the standard deviation of the population tuning function.

Eq. 2:

$$P(\bar{r}|s) = \prod_{i=1}^{N} e^{-f_i(s)} \cdot \frac{f_i(s)^{r_i}}{r_i!} \quad , \quad -\pi \leq s < \pi$$

$P(\bar{r}|s)$ is the actual population response activity given a stimulus, $r_i$ is the individual neuron contribution to the population response.

The integration (Layer III) is modeled as a weighted summation over different visual field locations where nearby locations weigh more than further ones. The weights are modeled as Gaussian functions described by Eq. 3 and shown in Figure 7.

Eq. 3

$$w_L(x) = \frac{1}{M\sigma_L\sqrt{2\pi}} \, e^{-\frac{1}{2}(\frac{x-L}{\sigma_L})^2}$$

$0 \leq x < 15$ : all possible visual field locations (deg)

$L = \{5, 7, 9\}$ : stimuli locations

$M_{SA} = M \odot \{2, \frac{1}{2}, 2\}$, $M_{FBA} = M \odot \{\frac{2}{3}, \frac{2}{3}, \frac{2}{3}\}$ : modulation factors

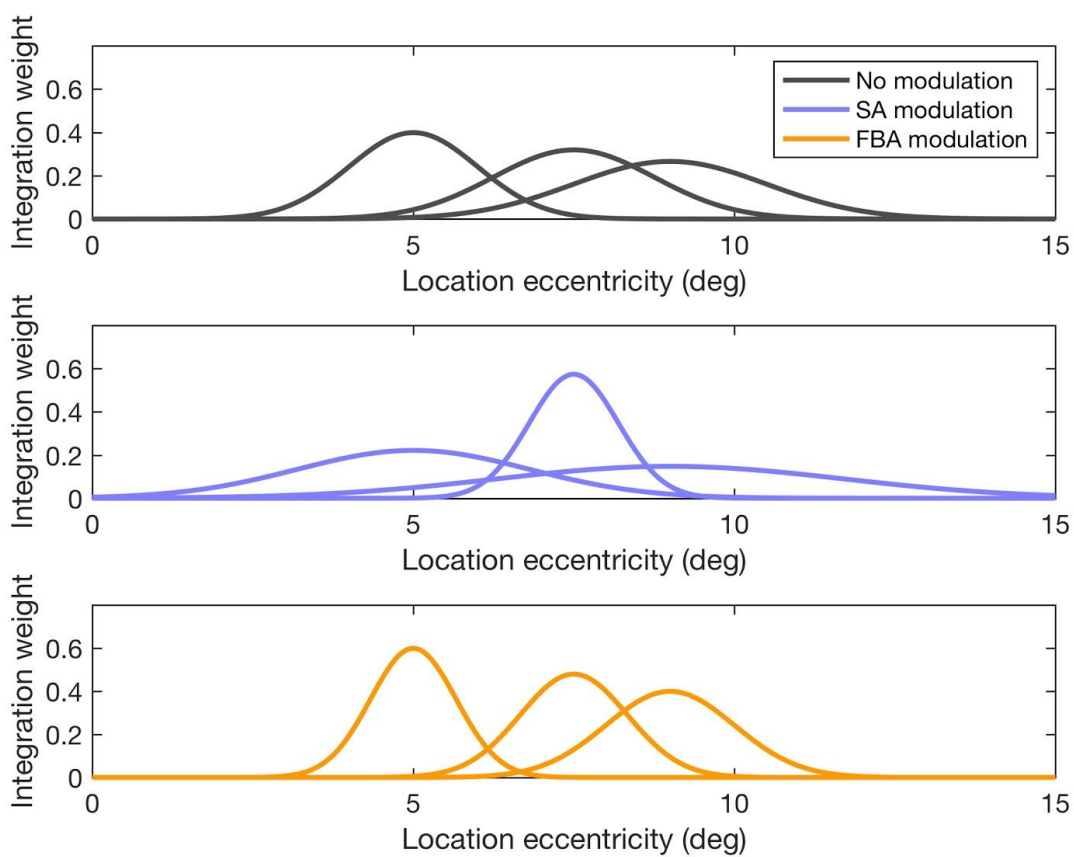

**Figure 7**. Examples of Gaussian weighting functions for the population responses integrated across different visual field locations. The non-modulated integration corresponds to the regular crowding effect. The spatial-selective SA modulation sharpens the weighting function at the attended location while broadening the others. The global FBA modulation sharpens (in a milder way) the weighting functions across the whole visual field.

The stimulus locations (indicated as *L* with values of 5, 7 and 9 degrees of eccentricity in our example) correspond to the means of these Gaussian weighting functions. Their standard deviations scale linearly with a slope derived by the linear regression of integration strength in the *baseline* condition over eccentricity (Figure 4-A, *slope* = 1.365). The way different attentional deployments modulate integration operates at this integration stage by tuning the standard deviations of the weighting functions: the spatial attention modulation ($M_{SA}$) reduces the standard deviation of the weight function centered on the focus of attention (in our case the peak of the lognormal fit to the probability density function of detections) and increases those outside this focus; the feature-based attention modulation ($M_{FBA}$) reduces the standard deviation of all weight functions across the visual field (albeit more modestly). Non-modulated *M* is a vector of identical values. The resulting integrated response at each examined location is described by Eq. 4.

Eq. 4

$$I_{L(i)} = \Sigma \left[ P(\bar{r}|s_{\{L\}}) \cdot w_{L(i)}(\{L\}) \right]$$

Finally, the integrated response is decoded (Layer IV) by fitting the response of Layer III with a gaussian mixture model with *k* components where *k* = number of examined locations. A decoded response results in a crowded percept in case there are multiple peaks in the integrated response. The higher the activity for the erroneous orientations (i.e. those of the distractors) compared to the correct ones (i.e. those of the target) the stronger the crowding percept would be. In the example shown in Figure 8 the target is the stimulus presented at 7 degrees with a pair of distractors at 5 and 9 degrees each.

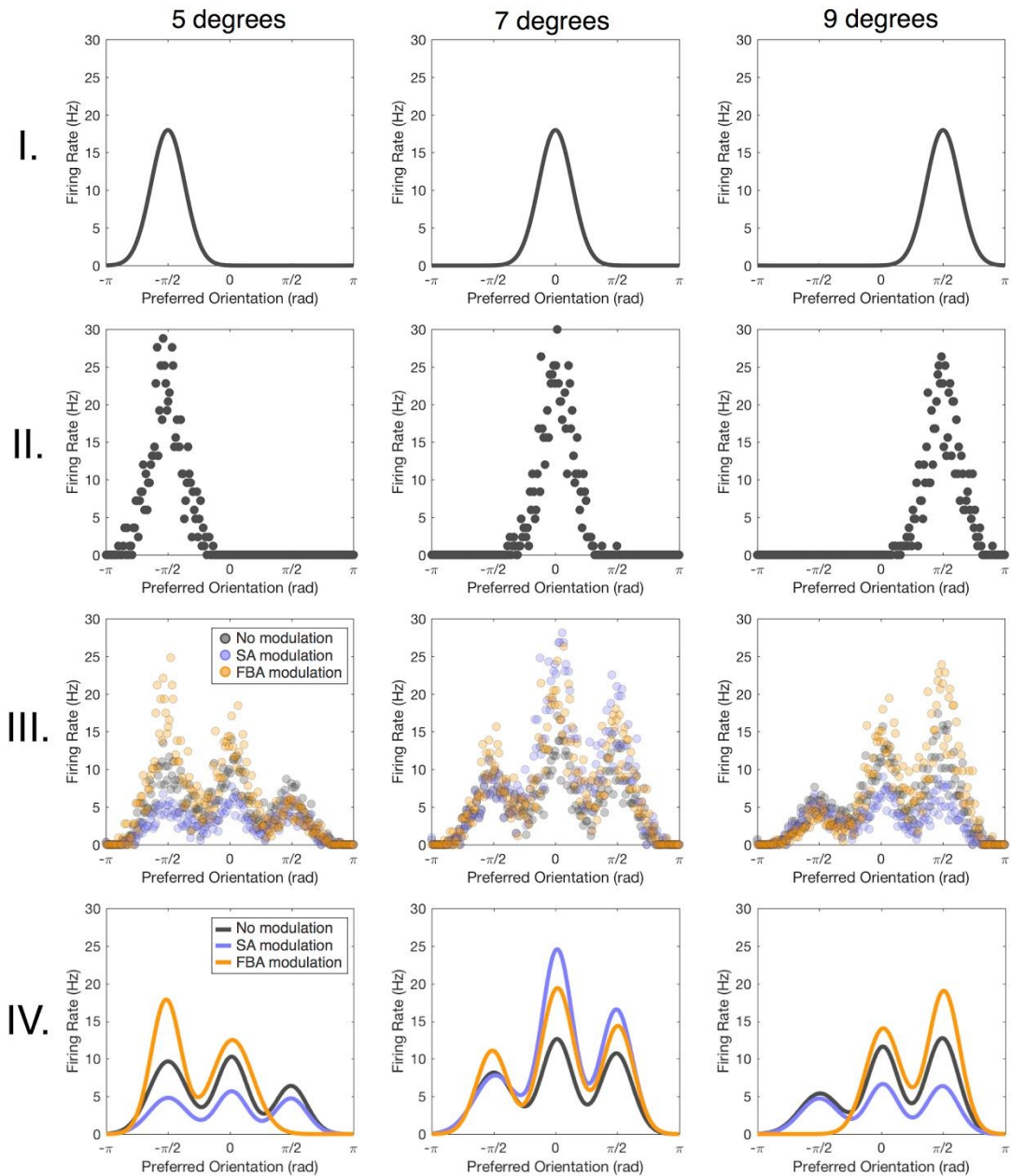

**Figure 8.** Graphical representation, layer by layer, of the population coding model with three sample locations (5, 7, 9 degrees of eccentricity) and three sample preferred orientations (-π/2, 0, π/2). **Layer I:** average firing rate of a neural population at the presentation of stimuli in different visual field locations; **Layer II:** neural activity (including noise) of the encoded stimuli; **Layer III**: integrated responses. Each location is integrated with nearby locations. The integration is computed as a gaussian-shaped weighted summation such that responses to nearby

stimuli receive more weight than farther stimuli. **Layer IV**: encoded population response probability distributions. The distributions are obtained by fitting the integrated responses with mixture gaussian models where each component represents a perceived orientation at that specific location. The attentional modulation that we propose acts by sharpening/broadening the weighting functions at the integration stage in Layer III, thereby changing the contributions that individual neurons make to the population response.

To quantify the average amount of information encoded by the neural population on a trial, we estimate the neural noise correlation in the integrated responses in Layer III. (Pouget et al., 2000). We kept fixed all model's input parameters (visual field locations of the stimuli, their orientations, response gain, and modulator functions described in Eq. 3 and Figure 7) so that the only variable parameter is the intrinsic neural noise that we want to measure (Eq. 2). We simulated 100 trials per each condition (*baseline*, *visual deprivation*, *information deprivation*) where we recorded the integrated activity of three neural populations at three different visual field locations (5, 7 and 9 degrees) being exposed to differently oriented stimuli. From this dataset we sampled the activity (expressed as firing rate, in Hz) of one random cell from each population (Figure 9, left side). To compute the correlation we treated the triplets of activity coming from the three population as coordinates in a 3-dimensional space (Figure 9, right side).

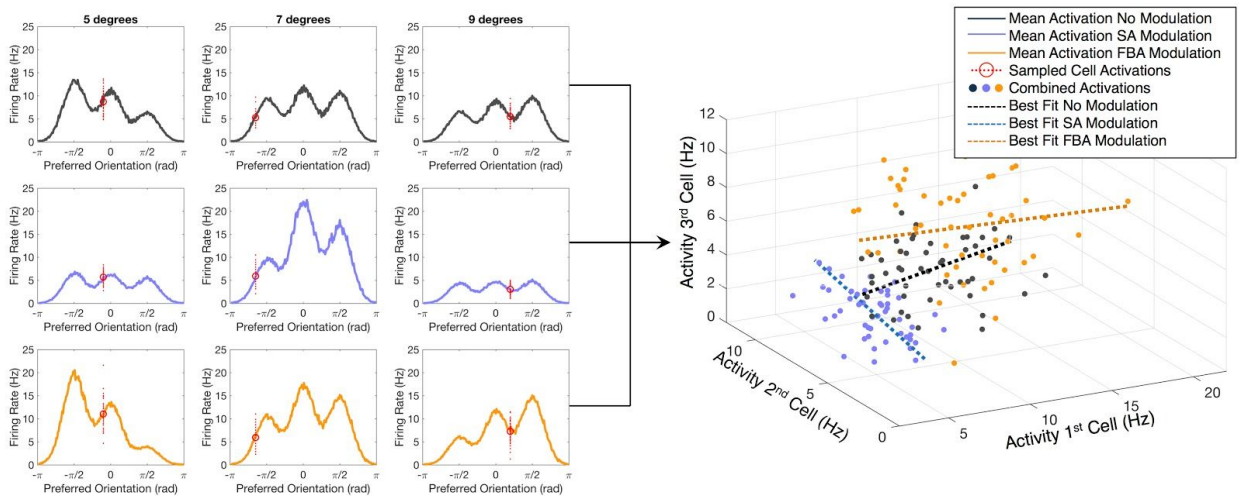

**Figure 9.** Model methods. Left side shows the integrated populations activity resulting from the simulation of 100 trials, where each trial is the presentation of three differently oriented stimuli at three locations in the peripheral visual field (5, 7, 9 degrees). From this pool of activities, we sample the trial-by-trial activities of one random cell (red dots) from each population. For every sample we compute the level of de-correlation between populations by means of a 3D orthogonal linear fit. The resulting triplets of activities from a single sample are shown on the right side. We repeat this sampling process 1000 times in order to compute the mean level of decorrelation between the neural populations at all tested locations.

Finally, the level of de-correlation was measured as the sum of residuals obtained from a orthogonal linear regression in 3D-space using Principal Component Analysis (Petras & Podlubny, 2006). The more the activity between neurons is de-correlated, the more information is carried by the examined population (Cohen & Kohn, 2011). Results of the model's outcome are shown in Figure 10.

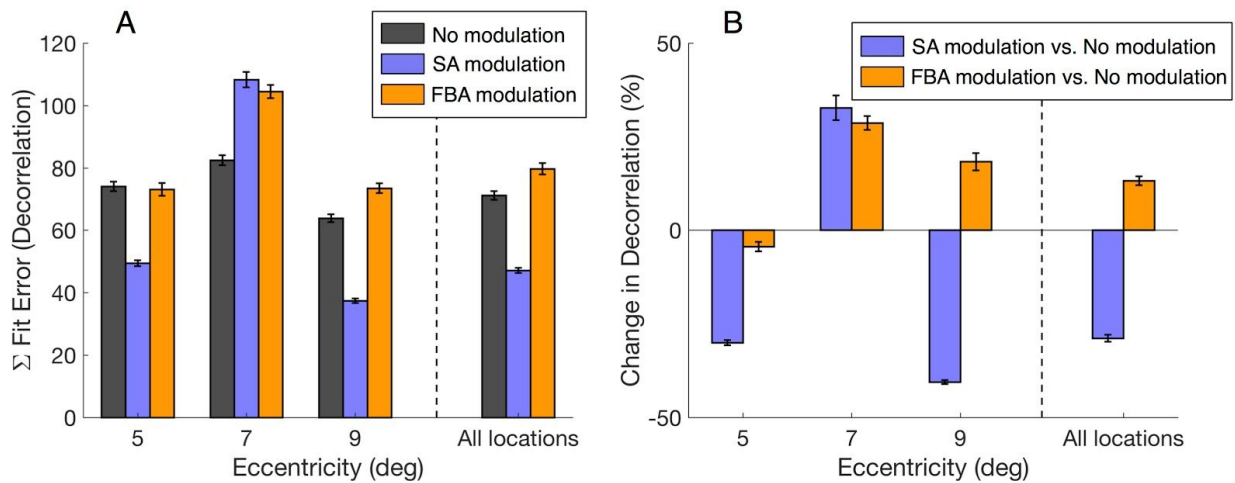

**Figure 10.** Model results. **A**, Levels of decorrelation at the three tested locations (with SA focused at 7 degrees). **B**, percentage changes in decorrelation of the *visual deprivation* and *information deprivation* conditions compared to *baseline*. De-correlation is expressed as the total fit error (sum of orthogonal residuals) resulting from a 3D orthogonal linear fit between the activities of a random distribution of triplet of cells, each one of them sampled from a different location. The decorrelations found by the model closely resemble the changes in integration strength found behaviorally: a marked increase correspondent at the location where SA is focused and a milder overall increase throughout the whole visual field when FBA is deployed.

In terms of information carried (expressed as de-correlation) we observe the two main effects shown by the behavioral experiments results: a marked increase in de-correlation at the location where spatial attention is focused and an overall, milder de-correlation through the whole visual field when the feature-based attention modulation occurs.

# Discussion

The main contribution of this study is a unified and mechanistic account of how spatial attention (SA) and feature-based attention (FBA) modulate spatial integration in the visual system. Our proposed model explains how attention modulates performance in crowded conditions, while leaving it untouched for isolated targets: sustained attention (be it either spatial or feature-based in nature) modulates the integration weights of the neurons in a population, while leaving their response properties intact. Consequently, our study shows how attention and integration interact to adapt and regulate the information flow through the visual brain. We will discuss our account and findings in more detail below.

### *Changes in visual integration strength are specifically related to attention*

We postulated that the *visual* and *information deprivation* conditions would induce a modulation in the deployment of SA and FBA, respectively, and we tested whether these changes affected visual spatial integration. The spatially non-selective nature of FBA (Boynton, 2005; Serences & Boynton, 2007) is consistent with the more mild and global reduction in integration strength that we observed in the *information deprivation* (i.e. removal of only task-relevant information: target orientation) condition. In contrast, in the *visual deprivation* (i.e. removal of all visual input) condition the reduction in integration strength was spatially selective, again being consistent with the nature of SA. At first sight these distinct findings would suggest separate underlying causes not necessarily related to attention. We will discuss these possibilities and why we concluded that a common attention-based explanation can account for the effects we observed.

First, the smoothed edge of the *visual deprivation*, as well as the complete removal of foveal stimulation, might have acted as a mask causing a local contrast adaptation (Ross & Speed, 1991). However, this would imply a change in orientation sensitivity also when the targets were presented in isolation – which we did not observe (Figure 3).

Second, the visible edge of the scotoma may have acted as a spatial cue. However, previous studies have concluded that spatial cueing does not alter crowding (Nazir, 1992; Scolari, Kohnen, Barton, & Awh, 2007; Wilkinson, Wilson, & Ellemberg, 1997). Moreover, evident from the identical pupil responses during the 2-AFC task, we showed that there was no significant change in attentional deployment, thus ruling-out a cueing-based explanation. In contrast to cueing, previous studies did find effects of sustained spatial attention on crowding (e.g. (Chen et al., 2014; He, Cavanagh, & Intriligator, 1996b; Intriligator & Cavanagh, 2001). Furthermore, the preferential retinal locus adopted by observers during the search task did not differ between *visual* and *information deprivation* condition (Figure 4-B), indicating that spatial cueing based on the presence of an edge cannot be a possible explanation for the attentional shift.

Finally, a well-established physiological index to investigate attentional deployment is pupil dilation: under cognitive demanding tasks the diameter tends to increase (Kahneman & Beatty, 1966)). Systematic changes in pupil size have been observed also for FBA and covert SA (Binda, Pereverzeva, & Murray, 2013, 2014)). In our results we observed a clear increase in pupil diameter over time between different conditions (Figure 5-A). Further, the fact that local variations around target detection do not differ significantly between conditions is another proof that sustained attention, rather than spatial cueing, is underlying the modulation of integration. We found the overall increase in pupil diameter only for the

*information deprivation* condition, which is visually identical to the *baseline* thus ruling out any other low-level factors that could alter pupil size. In contrast, in the *visual deprivation* condition the average diameter is not significantly different from the *baseline* condition. This might reflect a more complex interaction in pupil dilation between cognitive factors and change in foveal stimulation due to the complete lack of visual input. We found no significant differences in pupil dilation during the actual performance of the 2-AFC (Figure 5-B) implying that the effect of the modulation lasted beyond the duration of the actual attentional deployment.

*Attention modulates the neural activity underlying visual integration*

While FBA and SA are fundamentally different (e.g. (Boynton, 2005; Jacobs, Renken, Aleman, & Cornelissen, 2012)), the model based on our experimental results suggests that a single underlying mechanism can explain their influence on spatial integration. Our account is based on the notion of "population coding", as recent studies have shown that visual attention plays a crucial role in determining neural population responses (Kanashiro et al., 2017; Rabinowitz et al., 2015). Most population coding models of crowding assume that a decision about the presence of a target is made on the basis of an integrated signal. Attention can selectively modulate the activity of neurons by changing their response gain to, for instance, luminance contrast (Reynolds & Chelazzi, 2004) and attended features (i.e. orientation or color) (Boynton, 2005; Martinez-Trujillo & Treue, 2004) or adjusting their receptive field size and position (Connor, Preddie, Gallant, & Van Essen, 1997; Desimone & Duncan, 1995; Womelsdorf, Anton-Erxleben, Pieper, & Treue, 2006).

Indeed, a SA-induced increase in the activity of neurons in the region surrounding the *visual deprivation* increases the target's signal relative to that of the inner (and perhaps also outer) distractors, thereby selectively reducing the integration of objects presented in that region. Similarly, FBA enhances the activity of neurons selective for a particular feature (in our case orientation), irrespective of their location, causing a more global decrease in integration strength, as we observed in the *information deprivation* condition.

However, FBA comes in two flavours. It can either affect one specific feature over another (i.e. orientation vs. contrast) but can also affect the perception of different feature levels (e.g. horizontal vs. vertical orientations). We found significantly different sensitivity depending on the orientation of the targets in the 2-AFC task, regardless of the deprivation condition (Figure 6-A). However, we did not find any significant difference when performance was compared to *baseline* (Figure 6-B). In other words, these orientation-related differences cannot be linked to attentional modulation. This supports that the FBA modulation actually occurred by facilitating one feature category over others, rather than facilitating specific within-feature-levels.

*Candidate neural mechanism*

Our behavioral work cannot directly identify neural mechanisms. Still, based on our proposed model, we may speculate about potential candidate neural mechanisms that can mediate the attention-induced modulations of visual integration. Previously, horizontal connections in early visual areas have been proposed as the candidate neural circuitry underlying crowding (and thus integration; see (Levi, 2008). Our present findings are in line with this notion. Moreover, based on our findings, we can now additionally postulate that these horizontal connections must be adaptive and that their strength is modulated through feedback connections that mediate the attentional influences. Consistent with these ideas, horizontal connections play a major role in the facilitatory and suppressive interactions between the centers and surrounds of classical receptive fields (Fitzpatrick, 2000). Moreover, a previous study showed that feedback connections from later areas play a large role in determining the neural activity in the areas surrounding the classical receptive field (Angelucci & Bullier, 2003). Furthermore, feedback connections modulate the effectiveness of horizontal connections during perceptual learning (Gilbert, Li, & Piech,

2009). This implies that, although their anatomical length may be fixed, the strength of these horizontal connections can change, rendering the underlying the properties of the neuronal populations adaptive, as we witnessed in our present work. Hence, adaptive horizontal connections could underlie the modulation of integrated responses.

*Limitations and future studies*

Future studies could expand on our present work, by measuring the consequences of attention in further detail. For example, in our present experiment, we did not assess potential changes in the spatial extent of integration (i.e. critical distance, e.g. by changing the distance between target and distractors): according to our model, attention induces a narrowing of the spatial weighting function that determines the contribution of individual neuronal to the integrated signal. Consequently, we predict that critical distance – a classic measure in crowding research – would decrease under the influence of sustained attention (Chen et al., 2014; He et al., 1996a; Intriligator & Cavanagh, 2001). Moreover, we predict that the "uncrowded window" (Denis G. Pelli & Tillman, 2008) would simultaneously increase in size.

Neither did we vary the size of the scotoma: we predict that the locus of preferential detection and the locus of integration reduction would both show a shift in eccentricity that corresponds with the radius of the scotoma. This could help to further unravel the connection between integration due to crowding and ocular behavior (see also (Bedell, Siderov, Formankiewicz, Waugh, & Aydin, 2015; Nandy & Tjan, 2012). We would predict no or only little effect of changing the size of the information deprivation.

As indicated, our behavioral work also suggests a plausible neural substrate for the modulation in integration. fMRI or electrophysiological studies would be required to establish how attention modulates the population receptive fields (Klein, Harvey, & Dumoulin, 2014), amplitude of the stimulus representation (Sprague & Serences, 2013) or connectivity along the visual hierarchy. Moreover, it would be interesting to verify whether our model of attentional modulation of population weights can be generalized and also explain other attentional phenomena in visual perception. Given sufficient variation in the response properties of the neuronal population, changing integration weights in the feature-level domain could also account for a change in the average tuning of the neuronal population.

*Conclusion*

Attention can modulate visual spatial integration. It does so by adjusting the weights of individual neuronal responses, thereby increasing or decreasing their contribution to the integrated population response. We find two distinct modulatory effects of sustained attention on visual integration. Spatial attention modulation resulted in a spatially-selective reduction in integration strength, while feature-based attention modulation induced a more modest global reduction. Despite these distinctive effects, a single mechanism – adjusting integration weights at the population level – can coherently explain both. We propose that this mechanism provides the visual system with a flexible mechanism to optimize the processing of incoming visual information for the task it may have at hand.

## Competing interests

All the authors declare no competing financial and non-financial interests.

## Acknowledgements

This project has received funding from the European Union's Horizon 2020 research and innovation programme under the Marie Sklodowska-Curie grant agreement No 641805. We are thankful to Nomdo Jansonius, Joana Carvalho and Ronald van den Berg for their precious advice and support.